\documentclass[11pt]{article}
\usepackage{graphicx}
\usepackage{multirow}

\newcommand{\BABARPubYear}    {05}

\newcommand{\BABARConfNumber} {021}
\newcommand{\SLACPubNumber} {11374}

\input pubboard/babarsym

\def\KKKs      {\ensuremath{K^+ K^- \KS}\xspace}
\def\KKKz      {\ensuremath{K^+ K^- \Kz}\xspace}

\def\phiKs     {\ensuremath{\phi \KS}\xspace}

\def\mKK       {\ensuremath{m_{\Kp\Km}}\xspace}
\def\mKKs      {\ensuremath{m_{\Kp\KS}}\xspace}

\def\cosH      {\ensuremath{\cos \theta_H}\xspace}
\def\KKzz      {\ensuremath{X(1500)}\xspace}
\def\fz        {\ensuremath{f_0(980)}\xspace}
\def\qtag      {\ensuremath{q_{\mathrm{tag}}}\xspace}

\long\def\inst#1{\par\nobreak\kern 4pt\nobreak
    {\it #1}\par\vskip 10pt plus 3pt minus 3pt}

\begin{document}
{\pagestyle{empty}

\begin{flushright}
\babar-CONF-\BABARPubYear/\BABARConfNumber \\
SLAC-PUB-\SLACPubNumber \\
July 2005 \\
\end{flushright}

\par\vskip 5cm

\begin{center}
\Large \bf  Dalitz Plot Study of {\boldmath $\Bz \to \KKKs$}\ Decays
\end{center}
\bigskip

\begin{center}
\large The \babar\ Collaboration\\
\mbox{ }\\
\today
\end{center}
\bigskip \bigskip

\begin{center}
\large \bf Abstract
\end{center}
We present a study of the dynamics in $\Bz \to \KKKs$ decays
with approximately 230 million \BB~events collected by the
\babar~detector at SLAC. We find that the Dalitz plot distribution is best
parameterized with the \phiKs mode, an S-wave $\Kp \Km$ resonance near
1500~\mevcc, and a large non-resonant contribution. We set limits on
resonances not included in our model, and study models for the
non-resonant contribution.

\vfill
\begin{center}
Submitted at the 
International Europhysics Conference On High-Energy Physics (HEP 2005),
7/21---7/27/2005, Lisbon, Portugal
\end{center}

\vspace{1.0cm}
\begin{center}
{\em Stanford Linear Accelerator Center, Stanford University, 
Stanford, CA 94309} \\ \vspace{0.1cm}\hrule\vspace{0.1cm}
Work supported in part by Department of Energy contract DE-AC03-76SF00515.
\end{center}

\newpage
} 

\begin{center}
\small

The \babar\ Collaboration,
\bigskip

B.~Aubert,
R.~Barate,
D.~Boutigny,
F.~Couderc,
Y.~Karyotakis,
J.~P.~Lees,
V.~Poireau,
V.~Tisserand,
A.~Zghiche
\inst{Laboratoire de Physique des Particules, F-74941 Annecy-le-Vieux, France }
E.~Grauges
\inst{IFAE, Universitat Autonoma de Barcelona, E-08193 Bellaterra, Barcelona, Spain }
% ...

\end{center}\newpage

\section{INTRODUCTION}
\label{sec:Introduction}

Charmless three-body decays of the $B$ meson are a rich laboratory for
the physics of the Standard Model (SM), providing information on both
the weak sector and the dynamics of the strong interaction. Decays to
the final state \KKKs, which are dominated by $\b \to \s \sbar \s$
amplitudes, are of particular interest due to their sensitivity to
physics beyond the SM. However, little is known about the dynamical
structure of this decay.  Such an understanding is important because
although the branching fraction is
large~\cite{Garmash:2003er,Aubert:2004ta}, the final state is a
mixture of \CP-even and \CP-odd
states~\cite{Garmash:2003er,Aubert:2005ja}, hence any measurement of
the \CP\ asymmetry is diluted due to the presence of both
components. A key to more precise measurements of the \CP\ asymmetry
is to identify and parameterize various contributions to the final
state. An amplitude analysis, incorporating interference between decay
submodes, allows for the development of a model of the decay dynamics
and the extraction of the partial branching fractions and relative
phases for the modeled resonances.

In this paper we present results from a full amplitude analysis of the
decay $\Bz \to \KKKs$. The decay kinematics of a spin-zero particle
into three spin-zero daughters are completely determined by two
degrees of freedom. In terms of the invariant masses of daughter
pairs with four-momenta $p_i$, $m^2_{ij} = (p_i + p_j)^2$, the \Bz decay rate is
\begin{eqnarray}
	\frac{d\Gamma(\qtag, \mKK, \mKKs)} 
	{d\mKK^2 d\mKKs^2 }
	&=& \frac{1}{(2\pi)^3}\frac{1}{32 M_{\Bz}^2}  \frac{1}{2} \times 
\left [ \left | {\cal A} \right |^2 + \left | \bar{ {\cal A} } \right |^2   
 - \qtag  \frac{ \left | {\cal A} \right |^2 - \left | \bar{ {\cal A} } \right |^2 } {(\deltamd\tau)^2 +1}
	\right  ],
\label{eq::DP-q-Rate}
\end{eqnarray}
\noindent where $M_{\Bz}$, $\Delta m_d$, and $\tau$ are the mass, mixing
frequency, and lifetime of the \Bz, respectively. 
${\cal A}$ ($\bar{{\cal A}}$) is the signal \Bz (\Bzb) decay amplitude and \qtag is -1 (1) when the other $B$ meson in the event is a \Bzb (\Bz). Summing over \qtag, we obtain
\begin{equation}
	\frac{d\Gamma(\mKK, \mKKs)} 
	{d\mKK^2 d\mKKs^2} = \frac{1}{(2\pi)^3}\frac{1}{32 M_{\Bz}^2}\frac{1}{2}  \times \left [ \left | {\cal A} \right |^2 + \left | \bar{ {\cal A} } \right |^2  \right ].
\end{equation}
\noindent Assuming ${\cal A} \approx \bar{ {\cal A}}$, corresponding to no direct \CP asymmetry, this expression simplifies to
\begin{equation}
	\frac{d\Gamma(\mKK, \mKKs)} 
	{d\mKK^2 d\mKKs^2} \approx \frac{1}{(2\pi)^3}\frac{1}{32 M_{\Bz}^2}  \times \left | {\cal A} \right |^2.
\label{eq:decayrate}
\end{equation}
\noindent This assumption is consistent with the
cosine term observed in time-dependent \CP\ asymmetry measurements~\cite{Aubert:2004ta,Abe:2004xp}, and
predicted by preliminary calculations based on the QCD-factorization
model~\cite{Cheng:2005ug, Furman:2005xp}.

To describe the decay dynamics, we parameterize the amplitude in the
isobar model~\cite{isobar}, as a sum of contributions ${\cal A} =
\sum_r c_r \cdot {\cal A}_r$, where $c_r$ is a complex coefficient and
the index $r$ runs over the resonances in the model plus a
non-resonant component. For a resonance $r$ formed in the variable
$m^2_{ij}$,
\begin{equation}
{\cal A}_r(m^2_{ij},m^2_{ik}) = T_r(m^2_{ij}) \times F_B(|\vec{p_k}|) \times F_L(|\vec{p_i}|) \times Z_L(\vec{p_i},\vec{p_k}),
\end{equation}
where the momenta $\vec{p}$ are measured in the resonance rest frame,
$T_r$ is the dynamical function of the resonance, $F_L$ ($F_B$) is a
Blatt-Weisskopf barrier factor for the resonance ($B$ meson) decay,
and $Z_L$ is a Zemach tensor~\cite{BlattWeisskopf,Zemach:1963bc}.

The Blatt-Weisskopf factors $F_L(z)$, with $z = |\vec{p_i}| R$ and the
range $R = 1.5 \gev ^{-1}$~\cite{mesonradius}, depend on the resonance
angular momentum $L$~\cite{BlattWeisskopf}:
\begin{eqnarray}
F_{L = 0}(z) &=& 1, \nonumber \\
F_{L = 1}(z) &=& \sqrt{ \frac{1+ z_0^2}{1 + z^2}}, \\
F_{L = 2}(z) &=& \sqrt{ \frac{9+3z_0^2+z_0^4}{9+3z^2+z^4}}. \nonumber
\end{eqnarray}
Here $z_0 = z({|\vec{p_i}|}_0)$, where ${|\vec{p_i}|}_0=|\vec{p_i}|$ 
when $m_{ij} = m_r$, the resonance mass. For the factor $F_B$ for the parent $B$ meson, we use a range $R$ of zero.
The angular distribution of the decay
products for different $L$ is given in the Zemach tensor formalism by~\cite{Zemach:1963bc}:
\begin{eqnarray}
Z_{L = 0}(\vec{p_i},\vec{p_k}) &=& 1, \nonumber \\
Z_{L = 1}(\vec{p_i},\vec{p_k}) &=& -4 \vec{p_i} \cdot \vec{p_k}, \\
Z_{L = 2}(\vec{p_i},\vec{p_k}) &=& \frac{16}{3} \left [ 3(\vec{p_i} \cdot \vec{p_k})^2 - (|\vec{p_i}| |\vec{p_k}|)^2 \right ]. \nonumber
\end{eqnarray}

Most resonances are parameterized with the relativistic Breit-Wigner form
\begin{equation}
T_r(m^2_{ij}) = \frac{1}{m_r^2 - m^2_{ij} - i m_r \Gamma(m_{ij})}.
\label{eq:bw}
\end{equation}
$\Gamma(m_{ij})$ is the mass-dependent width, given by
\begin{equation}
\Gamma(m_{ij}) = \Gamma_r \left ( \frac{|\vec{p_i}|}{{|\vec{p_i}|}_0} \right )^{2L+1} \left ( \frac{m_r}{m_{ij}} \right ) \left ( F_L(|\vec{p_i}|) \right ) ^2,
\end{equation}
\noindent where $\Gamma_r$ is the nominal resonance width.
For the \fz, we use the coupled-channel
parameterization of Flatt\'{e}~\cite{Flatte:1976xv}, with couplings
taken from recent measurements~\cite{Ablikim:2004,Aitala:2000xt}.

The inclusive charmless branching fraction $\BR(\Bz \to \KKKz)$ is
measured to be $(24.7 \pm 2.3) \times
10^{-6}$~\cite{Garmash:2003er,Aubert:2004ta}, where
$(4.2^{+0.6}_{-0.5}) \times 10^{-6}$ is from the $\phi \Kz$
channel~\cite{Aubert:2003hz,Chen:2003jf,Briere:2001ue}. We have
previously reported on an angular moment analysis for this
decay~\cite{Aubert:2005ja}. As a function of \mKK, we found the $\phi$
to be the only statistically significant P-wave component. The S-wave
contribution is spread across the kinematic range, with a peaking
structure at around 1500~\mevcc. Contributions of higher waves were
consistent with zero. Our present analysis model is motivated by these
results. In addition to the $\phi(1020)
\KS$ channel, we include an S-wave resonance with a mass of about 1500~\mevcc, the \KKzz. An additional S-wave component, the
``non-resonant'' (NR) contribution, is added to account for the S-wave
events spread across the phase space. We also try adding the \fz to
the fit model.

The absence of higher waves in the \mKK angular analysis suggests the
lack of resonant channels decaying to $\Kp\KS$. The $\Bz \to
a_0^-(980) \Kp$ branching fraction has been measured as
small~\cite{Aubert:2004hs}. Additionally, there are theoretical
reasons to doubt the possibility of significant isovector
contributions~\cite{Laplace:2001qe,Chernyak:2001hs}.

Several decays involving $\b \to \c$ transitions can contribute to the \KKKs final state.
We coherently add an amplitude for the $\chi_{c0} \KS$ mode to our isobar model, while the
\Dp \Km and \Ds \Km channels are added incoherently and are described with
Gaussian shapes to account for the finite resolution in $m(\Kp\Km)$ mass.

Unfortunately, the present statistics do not allow for the parameterization of the P-wave outside of the $\phi\KS$ region
and the subsequent extraction of the relative $S-P$ phase.
Hence, we focus on the parameterization of the $S$-wave contribution which dominates the decay~\cite{Aubert:2005ja}.
Theoretical models of the NR amplitude~\cite{Cheng:2005ug,nr-theory} do not reproduce 
the distribution observed in data. 
Without reliable theoretical guidance, we test several ad-hoc parameterizations.
The three most successful models are:
\begin{eqnarray}
&&	\sqrt{\mKK^2- 4m_{\Kp}^2} ~ \left [ \mKK^2 \log \left (\frac{\mKK^2}{\beta_{nr}^2}   \right ) \right ]^{-1}~, \label{eq::nr_ccs}\\
&&	1~+~\beta_{nr} e^{ i \delta_{\beta}} \mKK^2  ,~\mathrm{and} \label{eq::nr_linear}\\
&&	\exp \left ( \beta_{nr} \mKK^2  \right ) ,\label{eq::nr_expo}
\end{eqnarray}
multiplied with isobar amplitudes and constant phases, and the
parameters $\beta_{nr}$ are free in the fits.
The first parameterization is inspired by a model from
Ref.~\cite{Cheng:2005ug}, where to improve agreement with data we have
dropped terms involving $\b \to \u$ transitions and those that give a
dependence on \mKKs. The second parameterization is a linearization of
theoretical models~\cite{nr-theory}, and the third model was successfully used in an
analysis of the decay $\Bp \to
\Kp\Kp\Km$~\cite{Garmash:2004wa}.  We choose the model of
Eq.~(\ref{eq::nr_expo}) to parameterize the NR amplitude. This model
gives good agreement with data and allows for direct comparison of the
$\beta_{nr}$ parameter with that determined in $\Bp
\to \Kp\Kp\Km$ decays.

We also test for a possible NR amplitude dependence on the $\Kpm\KS$ mass by adding
terms with a linear dependence on $\mKKs^2$.

Therefore, our main model includes four coherent
contributions: the $\phi(1020)$ ($m = 1019.456 \mevcc$, $\Gamma = 4.26 \mevcc$), \KKzz ($m = 1507 \mevcc$, $\Gamma = 109 \mevcc$), and $\chi_{c0}$ ($m = 3415.19 \mevcc$, $\Gamma = 10.1 \mevcc$) decaying to $\Kp\Km$; and
the NR S-wave~\cite{Eidelman:2004wy}.

\section{ANALYSIS METHOD}
\label{sec:Analysis}

This analysis uses a data sample of approximately 230 million \BB
events collected with the \babar\ detector~\cite{ref:babar} at the
SLAC \pep2\ \epem storage rings operating at the \FourS
resonance. The basic method used for reconstruction and selection of
\Bz-candidates is described in Ref.~\cite{Aubert:2004ta}. 
We characterize events with standard topological variables which
distinguish between the jet-like structure of the dominant continuum
\epem \to \qqbar ($q = u,d,s,c$) background and relatively
isotropic $B$ decays. These variables are combined into a Fisher
discriminant ${\cal F}$~\cite{Aubert:2004ta}. The shape of the ${\cal F}$ distribution for
continuum events is seen to vary as a function of the Dalitz plot
location, becoming more ``signal-like'' away from the Dalitz plot
edges. As a result, we do not include it in the fit, and instead
impose a requirement on the value of ${\cal F}$ to select events for the fit.

Two kinematic variables are used to characterize \Bz candidates, the energy
difference $\DeltaE = E_B - \sqrt{s}/2$ and the beam-energy-substituted mass $\mes =
\sqrt{(s/2 + \vec{p}_i \cdot \vec{p}_B)^2/E_i^2 - \vec{p}_B^2}$.
$E_B$ is the reconstructed energy of the \Bz candidate in the CM
frame and $\sqrt{s}$ is the total CM energy. 
$(\vec{p}_i, E_i)$ is the initial \epem four-momentum and $\vec{p}_B$
is the reconstructed \Bz-candidate momentum, both measured in the
laboratory frame. For signal events, \DeltaE peaks at zero with a
resolution of $18 \mev$, while \mes peaks at the \Bz mass with $2.6
\mevcc$ resolution. We initially retain candidates with $|\DeltaE| < 200
\mev$ and $\mes > 5.2 \gevcc$. For the Dalitz plot fit, we keep
candidates in a signal region (SR) of $|\DeltaE| < 60 \mev$ and $\mes
> 5.26 \gevcc$. We also define a sideband (SB) region, with $|\DeltaE| <
200 \mev$ and $5.2 < \mes < 5.26 \gevcc$, to study continuum
backgrounds. For calculation of the Dalitz plot variables, candidates
are refit with their mass fixed to the world average value~\cite{Eidelman:2004wy},
constraining the \Bz candidate to the kinematically-allowed Dalitz plot region.

Backgrounds due to $B$ decays are studied with samples of simulated
events. The largest contribution arises from the random combination of
tracks from both $B$ mesons.
This combinatorial background is difficult to differentiate
from continuum background, and is accounted for in the fits with the
continuum background. $B$ decays with a missing pion in the final state combination
fall outside both the SR and SB.

We use a high-statistics sample of simulated signal events to evaluate
the signal efficiency in bins on the Dalitz plot. The efficiency
varies primarily as a function of \mKK, falling from roughly 35\% for
low \mKK to about 10\% at high \mKK. The average efficiency is
26\%. Signal events which are misreconstructed with a daughter from
the other $B$ meson give a negligible contribution, making up less than
$0.5\%$ of the total events.

An unbinned extended maximum likelihood (ML) fit to the events in the
SR is used to extract event yields and amplitude
coefficients. Parameters of the probability density functions (PDF)
for signal \DeltaE and \mes are determined in simulated events, and
those for the continuum background are fit to data. For the amplitude
fit, we parameterize the event kinematics with the variables \mKK and
\cosH, where $\theta_H$ is the angle between the flight direction of the \Kp and \KS
in the $\Kp\Km$ center-of-mass (CM) frame. This change of variables
introduces a Jacobian term $|J| = (2 \mKK)(2 |\vec{p}_i| |\vec{p}_k| )$ in the signal PDF,
where $|\vec{p}_i|~(|\vec{p}_k|)$ is the \Kp (\KS) momentum in the \Kp\Km CM frame. The total PDF is formed as ${\cal P}(\mes)\cdot {\cal P}(\DeltaE) \cdot {\cal P}(\mathrm{Dalitz}),$ where for signal events
\begin{equation}
{\cal P}(\mathrm{Dalitz}) = d\Gamma(\mKK,\cosH)~|J|~\varepsilon(\mKK,\cosH)
\end{equation}         
\noindent before normalization, and $\varepsilon$ is the efficiency. The continuum ${\cal P}(\mathrm{Dalitz})$ PDF is modeled with a histogram of events from the SB region. Bin sizes are variable across the Dalitz plot to account for both narrow features and sparse statistics in different regions.

To explore the multiple solutions possible in the Dalitz fit, we
performed several hundred fits with the initial amplitudes and phases
randomized within reasonable values. In the fits, the amplitude
and phase of the NR contribution are fixed. The relative contributions of each component are
evaluated with the fit fraction $FF_r$, defined ignoring the interference
between amplitudes, as
\begin{equation}
FF_r = \frac{\int{|c_r {\cal A}_r|^2 |J| d\mKK d\cosH}}{\int{|\sum_s c_s {\cal A}_s|^2 |J| d\mKK d\cosH}}.
\label{eq::fraction}
\end{equation}

\section{SYSTEMATIC STUDIES}
\label{sec:Systematics}

We study several potential sources of systematic uncertainty. Fitting
to events generated with the PDF and with a full MC sample, we test
for any fit bias.  No significant biases are found, and we
conservatively include the largest deviation from the generated
parameters as systematic errors.  The parameters of the PDFs for the
kinematic variables are fixed in the fit. We vary them by one standard
deviation and take the change from the nominal fit as the systematic
error.  The masses and widths of all resonances except the \fz are
taken to be the world averages~\cite{Eidelman:2004wy}, and we vary the values by
their listed error and assign systematic errors based on the changes
in the fit results.  For the \fz, the systematic error due to the
parameterization is derived from the spread between the solutions
given by the coupling values measured by BES and E791~\cite{Ablikim:2004,Aitala:2000xt}.  We derive an analogous
uncertainty for the \KKzz by fitting with the parameters given by Belle for the
$f_X(1500)\to\Kp\Km$ observed in $\Bp \to \Kp\Kp\Km$
decays~\cite{Garmash:2004wa} and comparing the results with the
nominal fit that uses the world average values~\cite{Eidelman:2004wy}. We estimate
the systematic uncertainty due to the parameterization of the NR
component from the difference in results for the three models.

An additional contribution to the systematic error arises from the
spread of values at each of the quoted solutions. We assign the RMS of
this spread as the uncertainty. We observe that several of the
fit parameters are the same for multiple clusters of solutions, with
only one or two parameters differing. In these cases, we quote the
average as our result for the parameters that are nearly the same and
take the difference from the average as a systematic error on the
result.

\section{RESULTS}
\label{sec:Results}

\begin{figure}[!htb]
\begin{center}
\begin{tabular}{cc}
\includegraphics[width=3.25in]{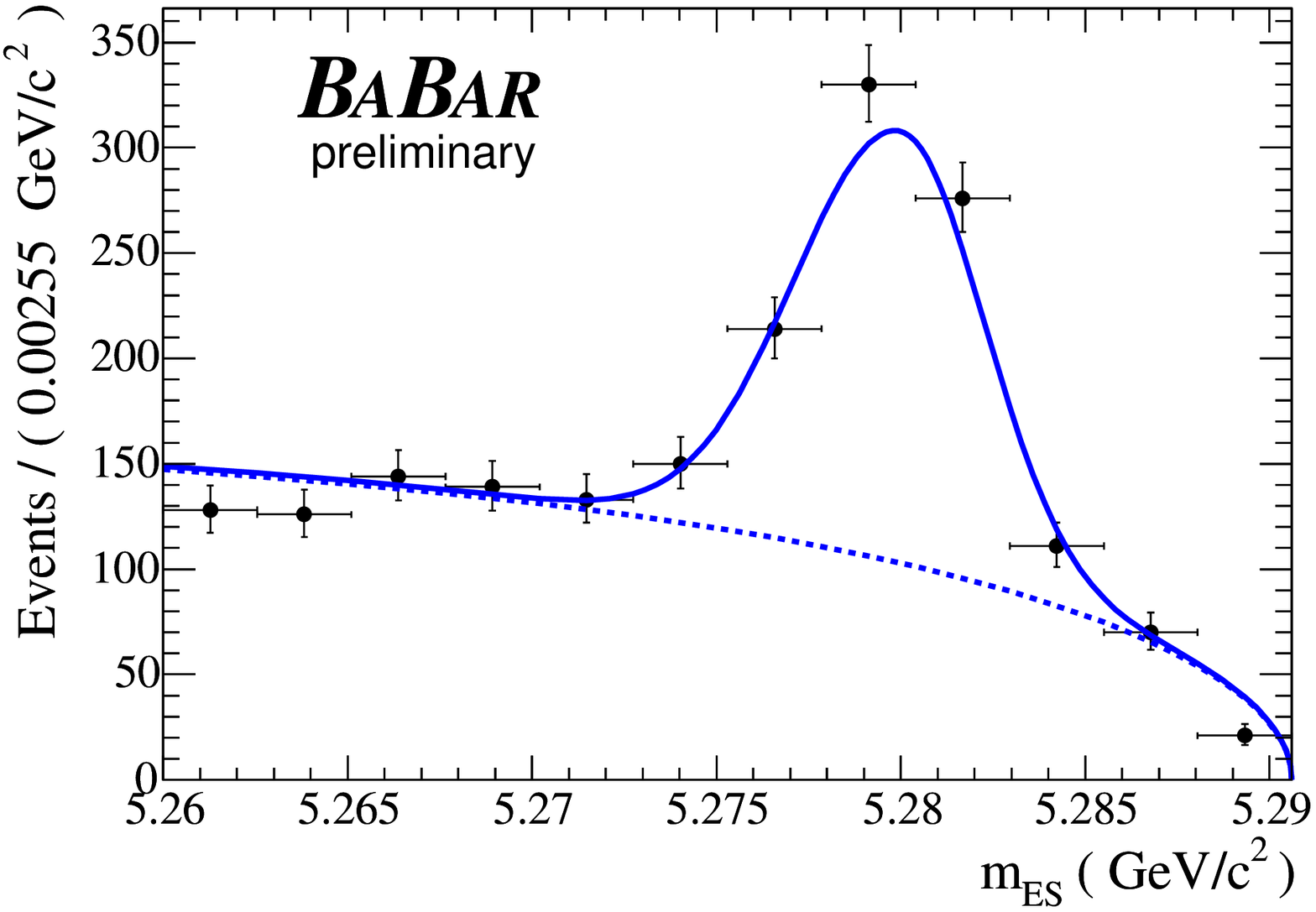} & \includegraphics[width=3.25in]{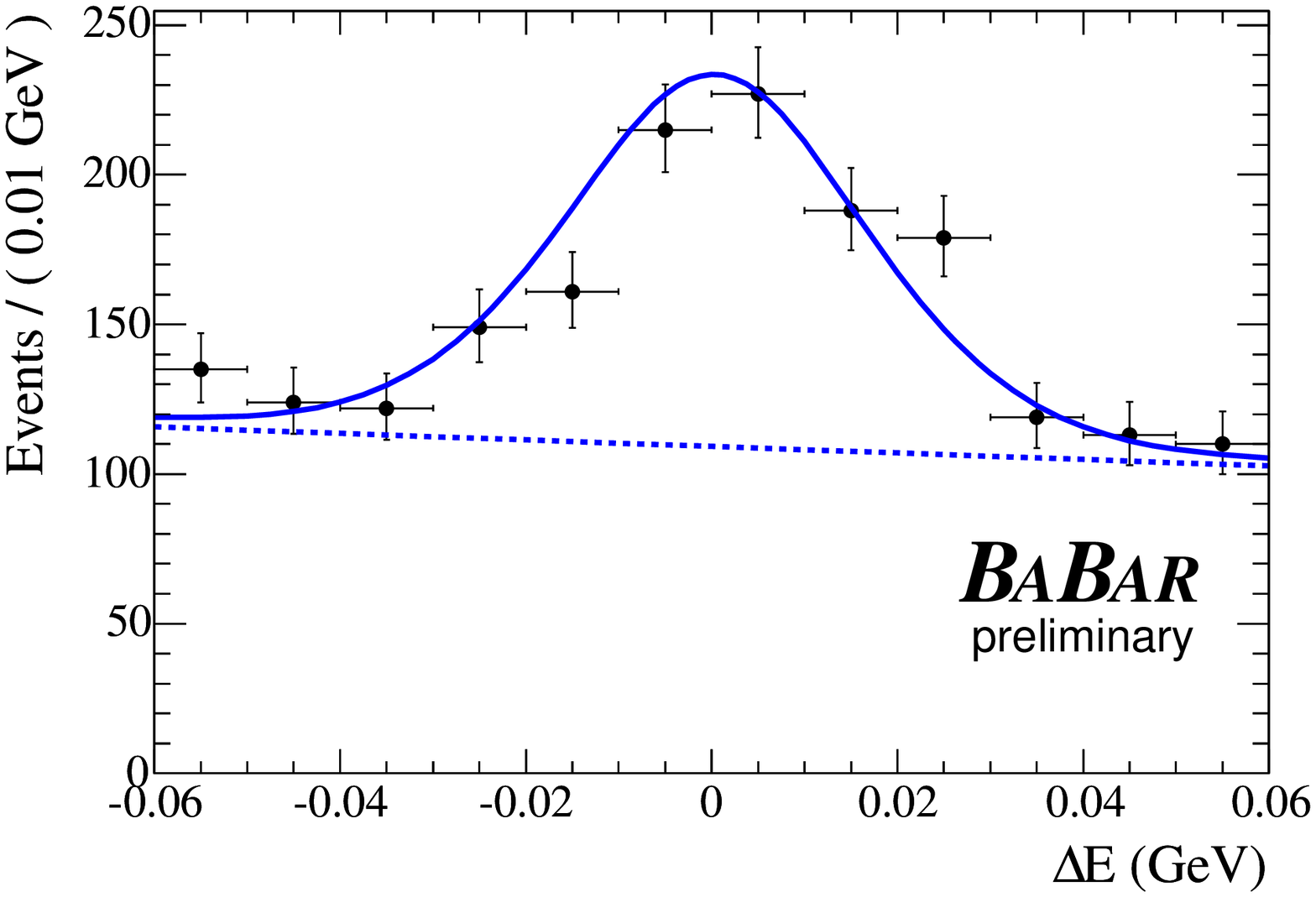}
\end{tabular}
\end{center}
\caption{Projections of the fit function for \mes (left) and \DeltaE (right) shown with data (points). The solid curve is the total PDF, and the continuum background PDF is shown as a dashed curve.}
\label{fig:kinplots}
\end{figure}

There are eleven free parameters in the fit: signal and background
yields, the NR parameter $\beta_{nr}$, and five strengths and three
relative phases. The phases for the three interfering components are
defined with respect to the NR component for which the phase of the
coupling to the $B$ meson is set to zero. In the extended ML fit with
a total of 1842 events, we find $530 \pm 28$ signal
events. Figure~\ref{fig:kinplots} shows the data overlayed on \mes and
\DeltaE projections of the fit function. Signal and background yields
are correlated with the Dalitz plot parameters at the 1\% level.

\begin{figure}[!htb]
\begin{center}
\begin{tabular}{cc}
\includegraphics[width=3.25in]{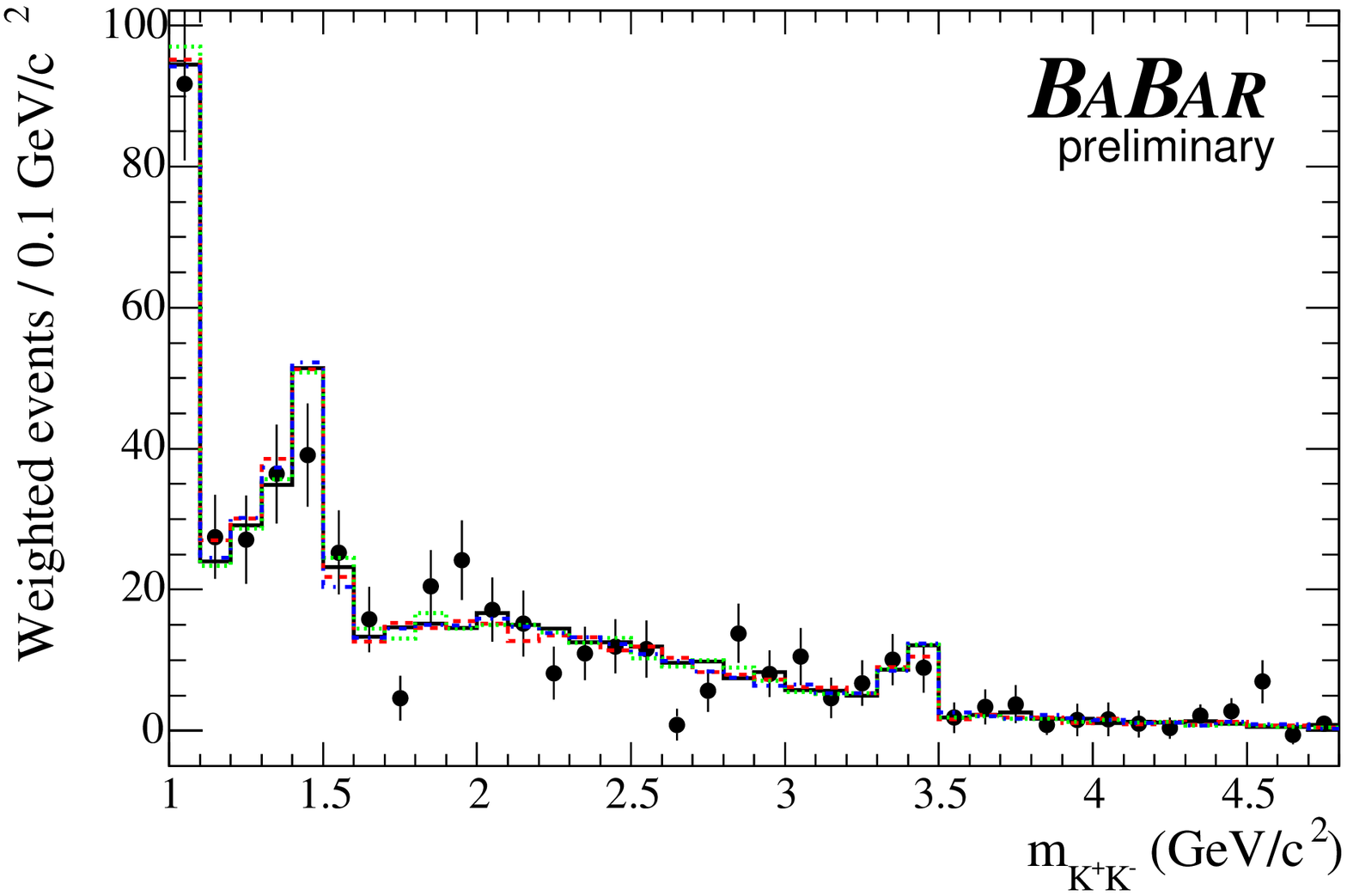} & \includegraphics[width=3.25in]{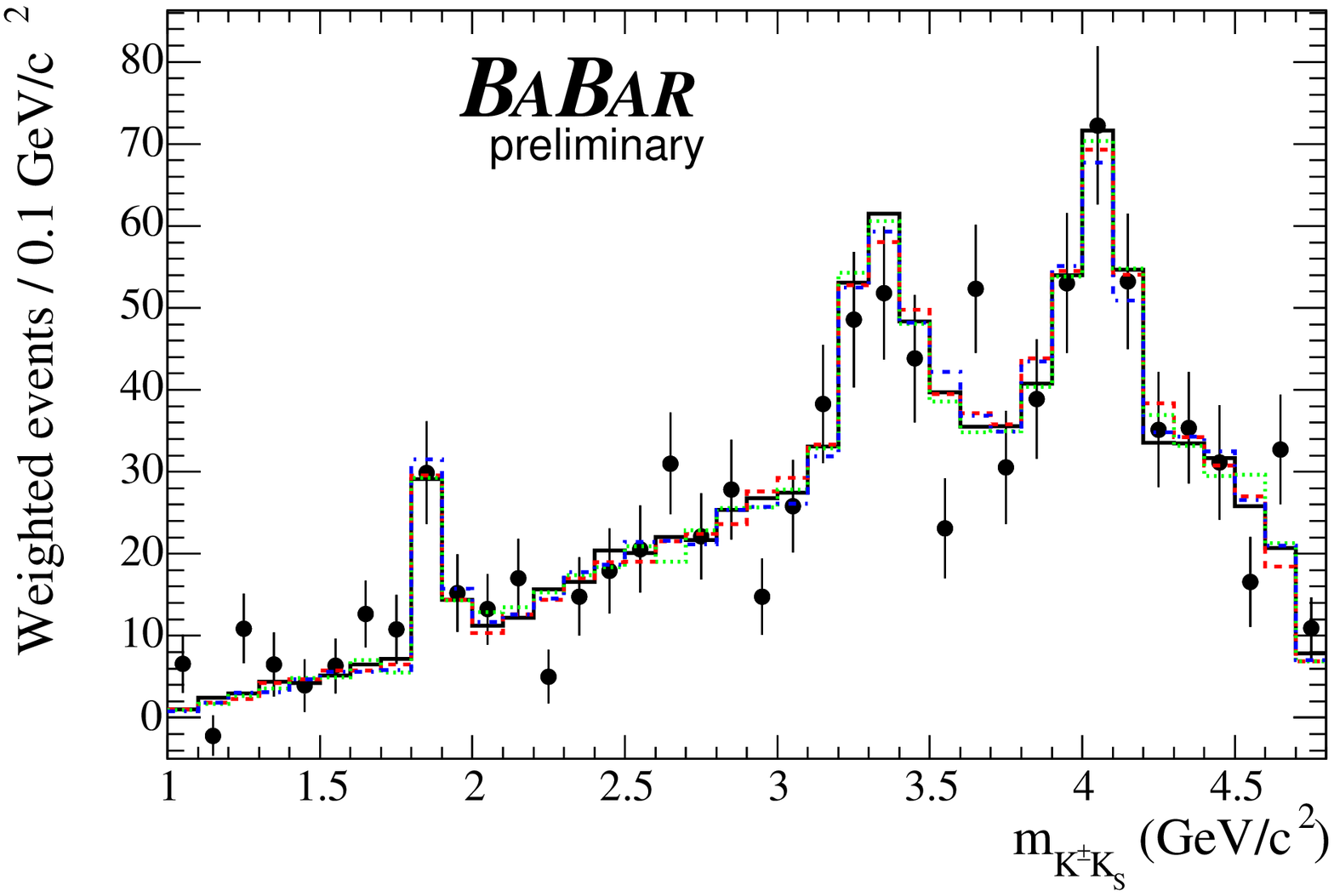} \\
\includegraphics[width=3.25in]{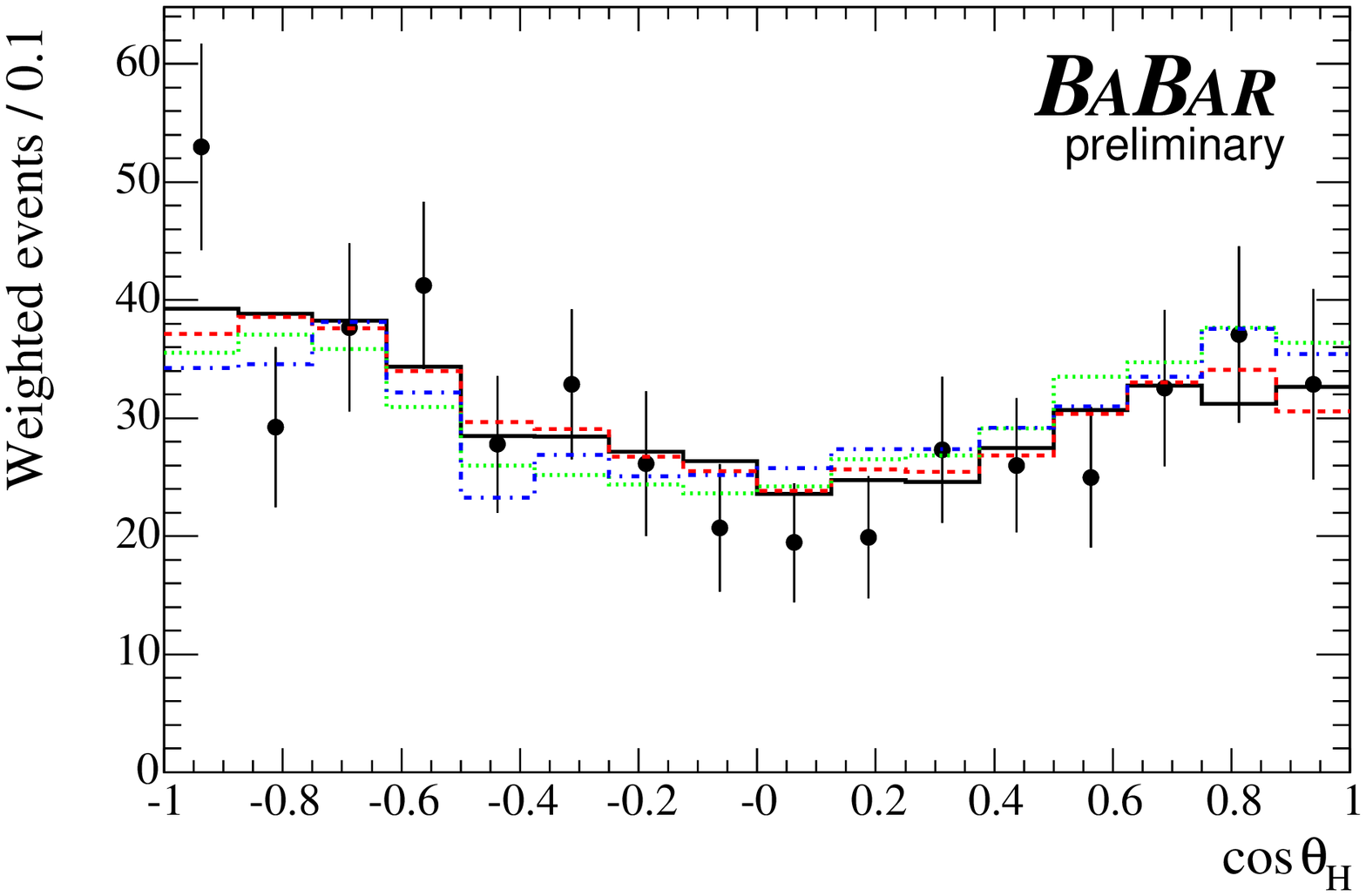} & \includegraphics[width=3.25in]{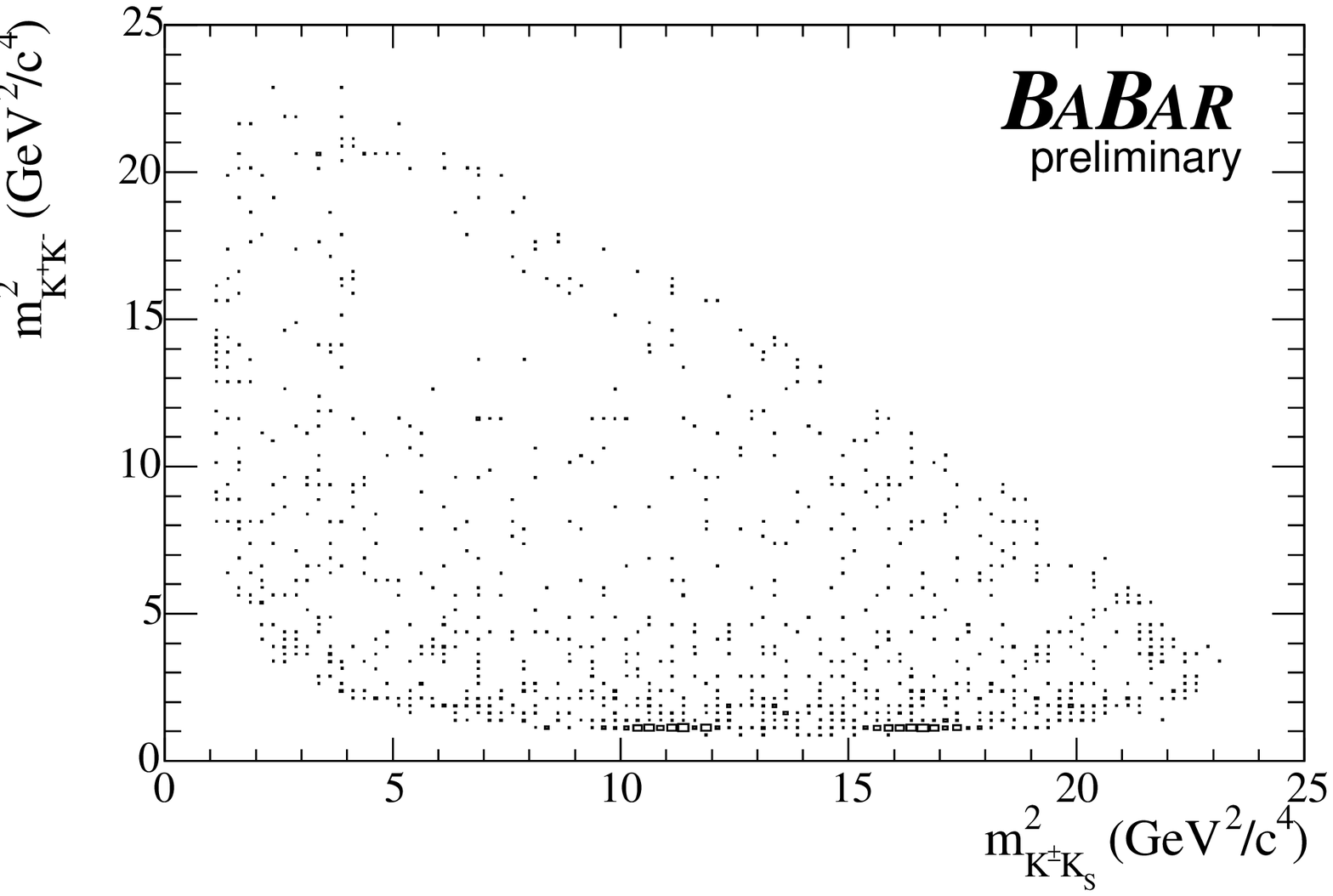}
\end{tabular}
\caption{Top row: \mKK (left) and \mKKs (right) projections of the nominal solutions (lines) with data (points).
For the \mKKs\ plot, there are two entries per event. The peak at $\mKKs \approx 1.8 \gevcc$ is due to $\Dp\Km$ and $\Ds\Km$ decays, while the peaks at higher values of \mKKs are reflections of forward and backward $\phi$ decays.
Bottom row: Left, \cosH projection of the nominal solutions (lines) with data (points); right, Dalitz plot showing distribution of signal events. Lines show solutions 1A (black solid), 1B (red dashed), 2A (green dotted), 2B (blue dash-dot) (See Tab.\ref{tab:solns}). All plots show background-subtracted signal events~\cite{Pivk:2004ty}.}
\label{fig:projplots}
\end{center}
\end{figure}

In several hundred fits with random initial values for parameters varied in the fit, we find two two-fold
ambiguities resulting in four solutions with similar likelihood values. 
We list these solutions in Table~\ref{tab:solns}, where the fit fraction $FF_r$ for each
component $r$ was defined in Eq.~(\ref{eq::fraction}).
Note that interference terms are neglected, so the sum of fit fractions does not necessarily 
add to 100\%. Plots showing projections of the fit function with data are shown in Figure~\ref{fig:projplots}.

We observe ambiguity in the fraction and the phase of $\KKzz \KS$
decays.  The solution with the relative phase close to zero has a
small value for the $\KKzz \KS$ fraction, but the fraction is 7
times larger when the phase is close to $\pi/2$. In
Tab.~\ref{tab:solns} and Fig.~\ref{fig:projplots}, we denote the
former solution ``A'' and the latter ``B''. Another ambiguity occurs
in the $\phi\KS$ component where there are two solutions approximately
3.5 radians apart in phase, but consistent in fraction. Here we use ``1''
and ``2'' to label the solutions. Both ambiguities are reproduced in
Monte Carlo studies where a fit to a sample of events generated
according to one of the solutions leads to the same ambiguities in the
fitted parameters as observed in the fit to data.

All NR models given by Eqs.~(\ref{eq::nr_ccs})-(\ref{eq::nr_expo}) result in a good fit, with
consistent fractions and phases.  
Using Eq.~(\ref{eq::nr_expo}), we get
$\beta_{nr}=-0.14\pm 0.02$, which is consistent with the parameter from the linearized model
of Eq.~(\ref{eq::nr_linear}). 
In the alternative model described with Eq.~(\ref{eq::nr_ccs}) we fix $\beta_{nr}=0.3$~\gevcc\ as 
suggested by~\cite{Cheng:2005ug}, and we get $\beta_{nr}=0.5 \pm 0.1 \gevcc$ if varied in the fit.
All fit solutions show the NR parameter to be inconsistent with a ``flat'' (phase-space) model. 
When we attempt to fit with the flat model, the fit does not converge in most cases (99\%), 
and we do not observe distinct clusters of solutions.  We also probe for a NR dependence on
the $\Kpm\KS$ mass by introducing an explicit dependence on $\Kpm\KS$ mass into our linear model, 
but the corresponding coefficients are consistent with zero. 

As a goodness-of-fit measure, we divide the phase space into bins and
compute the $\chi^2$ difference between signal-weighted
events~\cite{Pivk:2004ty} and the expectation based on the fitted
signal model. All bins have at least 10 events. We find for a
two-dimensional binning the value of $\chi^2/\mathrm{dof}=77.1/70$, and in
projections in \mKK and $\cos\theta_{H}$, respectively, we get
$\chi^2/\mathrm{dof}=48.9/40$ and $\chi^2/\mathrm{dof}=11.7/16$.  We find
negligible differences in $\chi^2/\mathrm{dof}$ among solutions.

\begin{table}[!htb]
\caption{Dalitz plot fit results. Labels used in the final column are described in the text. The upper limit for the $\fz\KS$ mode is at 90\% confidence.}
\begin{center}
\begin{tabular}{lrrrl}
\hline \hline
Mode				& $|c_r|$	&	Fit Fraction $FF_r$ (\%)		&	Phase		           & Solutions \\
\hline
\multirow{2}*{$\phi(1020)~\KS$}	& \multirow{2}*{$0.085 \pm 0.009$} & \multirow{2}*{$15.4 \pm 3.4 \pm 0.6$} & $-1.47 \pm 0.16 \pm 0.2$ & 1A, 1B	\\
				&		& 						&	$2.06 \pm 0.18 \pm 0.2$	   & 2A, 2B     \\
\multirow{2}*{$\KKzz$~\KS}      & $0.71 \pm 0.15$ & $5.2 \pm 2.2 \pm 0.9$     			&	$-0.17 \pm 0.21 \pm 0.1$   & 1A, 2A     \\
				& $1.93 \pm 0.18$ & $38.9 \pm 7.3 \pm 0.9$			&	$ 1.19 \pm 0.08 \pm 0.1$   & 1B, 2B     \\
Non-resonant			& $7.07$ (fixed) & $70.7 \pm 3.8 \pm 1.7$			&	$0$ (fixed)		   & All \\
$\chi_{c0}~\KS$			& $0.32 \pm 0.07$ & $3.1 \pm 1.6 \pm 0.8$			& 	$0.2 \pm 0.4 \pm 0.6$	   & All \\
\hline
$f_0(980)~\KS$	                & $1.57 \pm 0.41$ & $5.7 \pm 3.2 \pm 1.0$	($< 9.7$)		&	---	\\
\hline \hline
\end{tabular}
\end{center}
\label{tab:solns}
\end{table}

\section{DISCUSSION}
\label{sec:Discussion}

We find that the majority of the \Bz decays to \KKKs belong to the $\Kp\Km$ $S$-wave that we
identify as the non-resonant contribution. We find three equally good parameterizations that
describe this contribution as an $S$-wave with a dependence on $\Kp\Km$ mass.
When we apply the same model as in $\Bp \to \Kp\Kp\Km$ decays~\cite{Garmash:2004wa}, 
we find consistent values for the shape parameter.
We do not observe any variation of this component with $\Kpm\KS$ mass. This is consistent
with our previous angular moment analysis~\cite{Aubert:2005ja} that did not find
higher angular moments that would arise from such a variation.
We find good agreement with a theoretical model~\cite{Cheng:2005ug} only when 
ignoring amplitude terms with features that are not found in the data.

In addition to the non-resonant $S$-wave, we identify a
contribution with a $\Kp\Km$ mass around 1500~\mevcc. A two-fold ambiguity is
observed in the fraction, depending on the relative phase with respect
to the large non-resonant background.  The nature of this contribution
is unclear. Identification of this state as the $f_0(1500)$ is inconsistent
with the measurement of the $\Bz \to f_0(1500) \KS, f_0 \to \pip\pim$
decay~\cite{kellyford}. Since the ratio $\Gamma(f_0(1500) \to \Kp\Km)/\Gamma(f_0(1500)
\to \pip\pim)$ is $0.1845 \pm 0.0195$~\cite{Eidelman:2004wy}, we would expect only about two events in $\Bz
\to \KKzz \KS$. Our dataset has insufficient statistics for an independent determination of the mass and width of the state, so we use the world average values for the $f_0(1500)$~\cite{Eidelman:2004wy}.

In the low $\Kp\Km$ mass region, we find the $P$-wave contribution
from $\phi\KS$ decays with a fraction consistent with previous
measurements~\cite{Eidelman:2004wy}.  With the available statistics,
we do not observe a significant $S$-wave under the $\phi$
resonance. We perform a fit with the $f_0(980)$ resonance included in
the model and parameterized with the Flatt\'{e} lineshape, and set an
upper limit on the fraction for this decay of 9.7\% at the 90\%
confidence level.

\section{SUMMARY}
\label{sec:Summary}
We present the first study of the decay dynamics in $\Bz \to
\Kp\Km\KS$ decays and report preliminary results on the fractions and
relative phases of intermediate states that contribute to the
decay. This is of particular importance for the interpretation of
measurements of \CP~asymmetries in these decays, and searches for
physics beyond the Standard Model.  We assume no direct \CP\ asymmetry,
allowing us to use all events in this statistics-challenged
analysis. This assumption is supported by a cosine term in
time-dependent \CP\ asymmetry measurements consistent with zero, and
preliminary calculations based on the QCD-factorization
model~\cite{Cheng:2005ug, Furman:2005xp}.

One of the main goals of this paper has been to provide input to theoretical studies
that can result in better models. We extended our previous analysis~\cite{Aubert:2005ja}
by providing a parameterization for the dominant $S$-wave contribution, and
reported on features that are not seen in our dataset.
\section{ACKNOWLEDGMENTS}
\label{sec:Acknowledgments}

We are grateful for the 
extraordinary contributions of our \pep2\ colleagues in
achieving the excellent luminosity and machine conditions
that have made this work possible.
The success of this project also relies critically on the 
expertise and dedication of the computing organizations that 
support \babar.
The collaborating institutions wish to thank 
SLAC for its support and the kind hospitality extended to them. 
This work is supported by the
US Department of Energy
and National Science Foundation, the
Natural Sciences and Engineering Research Council (Canada),
Institute of High Energy Physics (China), the
Commissariat \`a l'Energie Atomique and
Institut National de Physique Nucl\'eaire et de Physique des Particules
(France), the
Bundesministerium f\"ur Bildung und Forschung and
Deutsche Forschungsgemeinschaft
(Germany), the
Istituto Nazionale di Fisica Nucleare (Italy),
the Foundation for Fundamental Research on Matter (The Netherlands),
the Research Council of Norway, the
Ministry of Science and Technology of the Russian Federation, and the
Particle Physics and Astronomy Research Council (United Kingdom). 
Individuals have received support from 
CONACyT (Mexico),
the A. P. Sloan Foundation, 
the Research Corporation,
and the Alexander von Humboldt Foundation.

\end{document}